\documentclass{mn2e}
\usepackage{graphicx}
\title {Is there a metallicity gradient in the LMC?}
\author[Feast, Abedigamba, Whitelock]{Michael W.
Feast$^{1,2}\footnote{Email: mwf@ast.uct.ac.za}$, Oyirwoth P.
Abedigamba$^{2,1}$ and
Patricia A Whitelock$^{2,1}$\\
$^{1}$ Astrophysics, Cosmology and Gravitation Centre,
Astronomy Dept, University of Cape Town, 7701 Rondebosch,
           South Africa.\\
$^{2}$ South African Astronomical Observatory, P.O.Box 9, 7935
           Observatory, South Africa.}

\begin{document}
\maketitle
\begin{abstract}
 A small but significant radial gradient in the mean periods
of LMC RR Lyrae variables is established from the OGLEIII survey data.
This is interpreted as a metallicity gradient but other possibilities are discussed.
Data on the ratio of photometrically selected C- and
M-type AGB stars in the LMC, kindly provided by M-R. L. Cioni, are
reanalysed.  Removing the effects of bias leads to conclusions
strikingly different to the original ones. 
There is a slight gradient of the C/M ratio in the inner part of
the LMC which might be due to a very small mean metallicity gradient. In the
outer part of the LMC the C/M ratio drops dramatically. The most likely
reason for this is that the proportion of older stars increases in the outer
regions. The mean metallicity of the inner AGB star population estimated
from the C/M ratio is lower than for intermediate age LMC clusters and
suggest that this population is in the mean older than the clusters and has
a mean age which falls in the LMC cluster age gap. 
\end{abstract}

\section{Introduction}
 To understand the origin and evolution of the LMC we need to know how
objects of different ages and metallicities are distributed within it. A step in
this direction is to establish whether there are metallicity gradients in
LMC objects of various types. Pagel et al. (1978) discussed the oxygen
abundance of LMC H{\sc II} regions and found marginal evidence for a decrease
outwards from the centre. More recently a radial metallicity gradient for
the AGB star population has been discussed by Cioni (2009)(=C09) who also
summarized other relevant work. In the present paper we report a search for
a possible metallicity gradient in the RR Lyrae population based on the LMC RR Lyraes
discovered by the OGLE group (OGLE-III, Soszy\'{n}ski et al. 2009 =
S09) together with a period-metallicity relation.  The RR Lyraes, as tracers
of the oldest populations, are particularly important for understanding the
early history of the LMC. We also compare our results with previous work
including a reanalysis of the metallicity gradient from AGB stars.

\section{The period-metallicity relation}
 A number of authors have discussed the relation between period and
metallicity for Galactic RRab Lyrae stars (fundamental mode RR Lyraes). Fig.~1 
is a plot of such data, kindly provided by Dr A. Layden. Also shown in
the figure are data for the LMC. For these, the spectroscopic values of
[Fe/H] derived by Gratton et al. (2004) and Borissova et al. (2006) were
used. Two lines are shown in the figure. The dotted line is a least square
fit to the Layden data. This has the relation:
 \begin{equation}
[Fe/H] = -5.62 (\pm 0.47) \log P_{ab} -2.81 (\pm 0.13), \sigma = 0.42.
 \end{equation}
The solid line is given by:
\begin{equation}
[Fe/H] = -7.82 \log P_{ab} -3.43, \sigma = 0.45,
\end{equation} 
 which is the relation used by Sarajedini et al. (2006) for an investigation
similar to the present one, but on RR Lyraes in M33. 
\begin{figure}
\includegraphics[width=8.5cm]{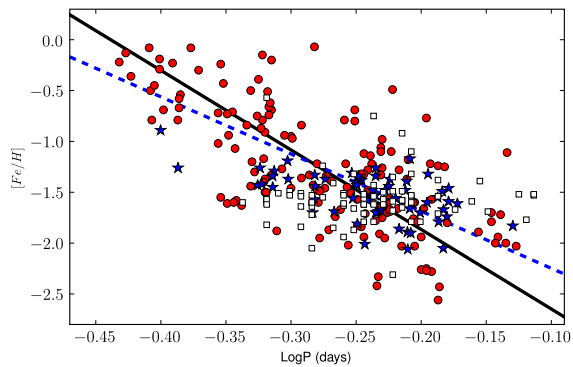}
\caption{The relation between [Fe/H] and $\log P_{ab}$ for Galactic RR
Lyraes (circles from Layden) and LMC RR Lyraes (stars from Gratton and
squares from Borissova) the lines are for equation 1 (dashed) and equation 2 (solid).}
\end{figure}
 It is based on a slightly different sample of Galactic RR Lyraes which was
also supplied to them by Dr A. Layden.

Evidently the Galactic data show a clear trend though with considerable scatter.
The LMC data generally lie together with the Galactic points.
We shall assume that LMC RR Lyraes follow the Galactic relation and discuss below
the effect of possible deviations from it. 

Suggestions have been made that a relation involving both period and light
curve shape gives an improved method for estimating metallicity (e.g.
Jurcsik \& Kov\'{a}cs 1996). We tested this for the available LMC
light-curve parameters (S09) and found no improvement in that case. Our
analysis is therefore based on equations 1 and 2.

\section{The LMC metallicity gradient for RRab variables}
 The LMC OGLE-III catalogue lists 17,693 RRab stars. Most of these are
confined to a limited magnitude range (e.g. S09 fig.~7). There are, however,
a small number at brighter magnitudes. These are likely to include blended
stars and foreground objects. We have therefore omitted all stars with
$I<18.3$ mag. Tests show that including or omitting these data does not
affect our results in any significant way. We have also omitted all RR
Lyraes marked by S09 as blended or questionable and also those in the region
of LMC globular clusters as defined by S09. This left us with a sample of
16,864 RRab stars.

 In her discussion of the the metallicity distribution of AGB stars, Cioni
(C09) adopts a model for the LMC which is essentially that of a disc
inclined to the plane of the sky.
The true three dimensional distribution of RR Lyraes in the LMC is not at
present known and, as a first approximation, we follow the model used by
Cioni. This adopts an inclination of $34^{\circ}.7$, a position angle of the
major axis of $189^{\circ}.3$ and an LMC distance of 51kpc. Changing the
distance will not significantly change any conclusions and Cioni's distance
is adopted so that a direct comparison can be made with her results. Using
this model we determine the distance $R_{GC}$ of a variable for the centre
of the LMC, taken (with Cioni) to be $\alpha = 82^{\circ}.25$ and $\delta =
-69^{\circ}.5$.

For each RRab star we derive an estimate of [Fe/H] from equations 1 or 2
and use these data to obtain mean [Fe/H] in concentric annuli about the
centre. The results are shown plotted in Fig.~2.  The errors are internal,
i.e. they not include the uncertainties in equations 1 and 2. Fig.~2 indicates a
mild systematic gradient outward and least square fits are shown. These are,
based on equation 1:
\begin{equation} [Fe/H] = -0.0104 (\pm 0.0021) R_{GC} -1.4213 (\pm 0.0046), 
\end{equation} 
and based on equation 2: 
\begin{equation} [Fe/H]=-0.0145 (\pm 0.0029) R_{GC} -1.4976 (\pm 0.0063). 
\end{equation}
 These two relations (equations 3 and 4) give nearly equal gradients. The
absolute values of the metallicities differ by less than the standard error
of the zero-point in equation 1
 \begin {figure}
\includegraphics[width=8.5cm]{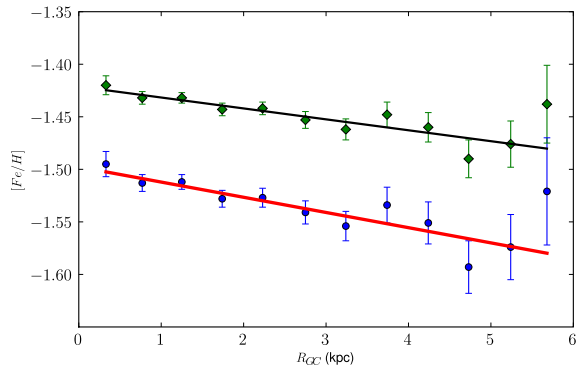}
\caption{The [Fe/H] - $R_{GC}$ relations for the RR Lyraes in the LMC. The
upper curve (diamonds) uses equation 1 and the lower curve (circles)
equation 2.}
\end{figure}


The longer period RRab stars are on average more metal poor and also of
lower light amplitude than the shorter period ones. One might therefore be
concerned that the above results could be affected by a lower efficiency of
finding low amplitude variables in the more crowded regions of the LMC. We
have made an attempt to check whether this is a significant effect by
determining the ratio of the number of RRab stars to RRc stars as a function
of distance, $R_{GC}$, from the centre. The overtone, RRc, variables are of
lower amplitude than corresponding RRab stars. Thus, other things being
equal, a selection effect should show as an increase of the $N(ab)/N(c)$
ratio (loss of RRc variables) in the inner parts. Fig.~3 shows no evidence of
this. If anything there may be a slight decrease in the ratio in the inner
parts, but this is quite uncertain. Of course the ratio may well depend on
metallicity, but at least in the Galactic globular cluster population this
works in the sense that the ratio is smaller in the more metal-poor
clusters.


\begin{figure}
\includegraphics[width=8.5cm]{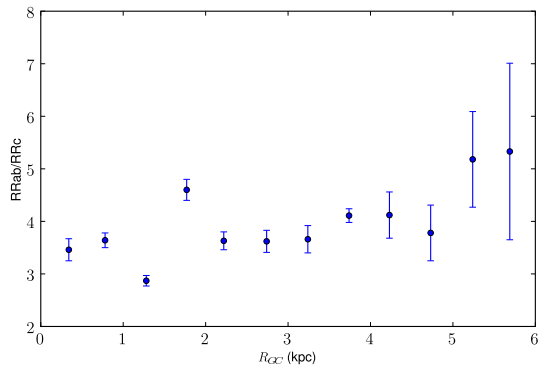}
\caption{The ratio of the numbers of $RR_{ab}$ to $RR_{c}$ stars as a function of
$R_{GC}$ in the LMC.}
\end{figure}

As already noted, Gratton et al. (2004) and Borissova et al. (2006) have
derived estimates of [Fe/H] in the LMC from low resolution spectra. Fig.~4
shows their results plotted against $R_{GC}$. Variables in the region of LMC
globular clusters are omitted.  A least squares fit to the data is shown,
this has the form:

\begin{equation}
[Fe/H] = -0.050 (\pm 0.037) R_{GC} -1.480 (\pm 0.045), \sigma =0.18.
\end{equation}

These data, which cover the inner region of the LMC agree with the results
from the period distribution within the uncertainties; though taken by
themselves they do not provide convincing evidence of a metallicity
gradient. 
\begin{figure} \includegraphics[width=8.5cm]{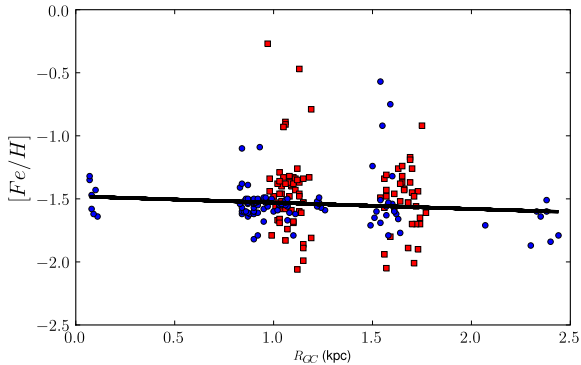}
\caption{ The spectroscopically determined metallicities of LMC RR Lyrae
variables plotted as a function of $R_{GC}$. The (red) squares are from
Gratton et al. (2004) and the (blue) circles from Borissova et al. (2006).}
\end{figure}

The results summarized in Fig.~2 depend on the assumption that the
LMC RR Lyraes follow the same mean period - [Fe/H] relation as the Galactic
stars. This assumption cannot be tested directly. The LMC RR Lyraes with
spectroscopic abundances plotted in Fig.~1 lie with the Galactic points. However,
their small number, their limited range in [Fe/H] and the substantial intrinsic
scatter in $\log P_{ab}$ at a given metallicity \footnote {A substantial range of
periods at a given metallicity is expected in view of the range of periods in a given
globular cluster.}, precludes any attempt  to derive an independent slope to the
relation. A referee has suggested that we should consider the case when there is
no relation between [Fe/H] and period and that there might in fact be no metallicity
gradient for this population. Since Fig.~2 is simply a scaled version of a mean
$log P$ - $R_{GC}$ relation, the gradient in mean $\log P$ must then be explained
in terms other than [Fe/H]. A model with a nearly constant mean
[Fe/H] for the RR Lyraes, but with an age gradient might be possible. It will be clear that our
favoured model of a metallicity gradient might well also imply an age gradient.
More complex models might be developed. However, they will all involve a systematic
radial gradient in RR Lyrae properties.


\section{A possible metallicity gradient for the AGB star population}

Battinelli \& Demers (2005) suggested that in a stellar system the ratio of
carbon-rich to oxygen-rich AGB stars was a function of [Fe/H] and a
rederivation of this relationship was given by C09 who found:

\begin{equation}
[Fe/H] = -0.47 (\pm 0.10) \log (C/M)- 1.39 (\pm 0.06),
\end{equation}
though with considerable scatter as can be seen from fig.~B1 of C09.

Cioni \& Habing (2003) selected likely LMC AGB stars (i.e. they were brighter
than an adopted tip of the RGB) from the DENIS survey in $IJK$, and divided
them into probable C and M stars using colour criteria. These data have been
used by C09 to study the C/M number ratio and to estimate [Fe/H] using
equation 6 as a function of distance from the centre of the LMC.
In this work,
the numbers of likely C and M stars were counted in bins of size 0.04
square degrees over a grid of $100\times 100$ bins. 

Dr Cioni has very kindly placed these counts at our disposal. We proceed as
in subsection 3 and calculate the value of $R_{GC}$ for each bin and then
sum all the likely C and M stars in the bins lying in annuli about the
centre.


The results are plotted in Fig.~5.  This figure shows a  very slight increase
in C/M from $R_{GC} =1$ to $R_{GC} = 4$ kpc. If this is taken as significant
it represents, using equation 6, a decrease in [Fe/H] from $\sim-1.07$ in the
centre to $\sim-1.10$ at $R_{GC} \sim 4$ kpc. It seems to us that the drop in
C/M at greater values of $R_{GC}$ can be hardly be attributed to an increase
in mean metallicity. It seems much more likely that this is due to a change
in age of the dominant population. It is natural to assume that this is due
to older populations being more prominent on the outskirts of the LMC. 
These, of course, might be metal poor. 

\begin{figure}
\includegraphics[width=8.5cm]{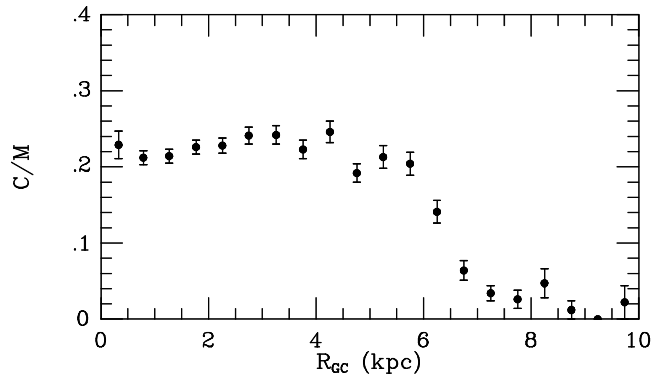} 
\caption{The AGB C/M ratio for
the colour-selected sample as a function of $R_{GC}$.} \end{figure}

These conclusions are surprisingly different from those in C09 and we are
very grateful to Dr Cioni for send us details of her work. This consisted of
deriving the C/M for each bin and these, converted to [Fe/H] using equation 6 are
plotted in C09 fig. 2. A problem arises in this case since in the outer parts
of the LMC the number of C or M stars per bin is very small and may be zero. 
For instance in the annulus at $R_{GC}= 3.76$kpc, there are 408 bins of
which 4 have no M stars, 96 have no C stars and two have neither C nor M
stars.
The procedure in C09 was to omit bins when either the number of likely C
stars or M stars was zero. Since in general $\rm C/M < 1$, this will tend to
bias the results towards higher C/M ratios.
 That this is a significant effect can be seen from Fig.~6 which shows results 
from counts in annuli as before, but now with all bins omitted which have no
likely C stars or M stars. Within the uncertainties and the question of
assigning proper weight to individual bins, Fig.~6 converted to [Fe/H] by
equation 6 would replicate the results of C09. Thus we believe these results
are significantly affected by this selection procedure.

\begin{figure}
\includegraphics[width=8.5cm]{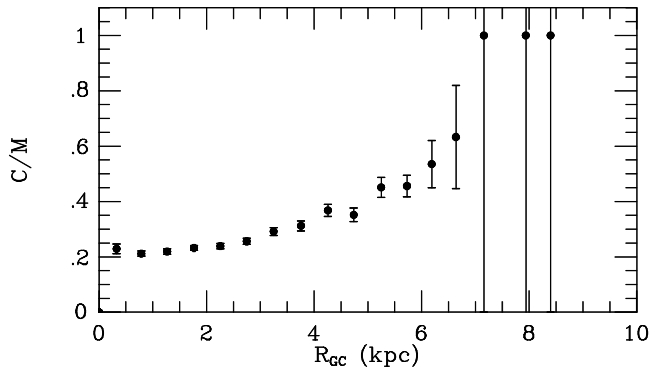}
\caption{The same as Fig.~5 except that bins with either no C or no M stars
were omitted. Note the compressed vertical scale compared with Fig~5.}
\end{figure}


It is at first sight rather strange than in an initial discussion of these
data Cioni \& Habing (2003) obtained a plot (their fig.~3) which appears to
show a ring of larger than average C/M values in the outer parts. This used
data from individual bins.  A possible explanation of this is as follows.
The apparent ring occurs in regions where the number of stars per bin is
small and thus the standard error of a C/M ratio is large. Thus even if C/M
were in the mean constant the scatter (which will be skew) increases as the
number of stars per bin decreases and leads to the occurrence of some large
ratios in the outer parts. It is possible that this, together with the very
slight apparent maximum in mean C/M at $R_{GC} \sim 4$ kpc in our work
(Fig.~5), is enough to produce the apparent effect seen in the Cioni \& Habing
figure.

During a study of the C09 data we found that, whilst the distribution of C
stars among the bins of a given annulus was apparently random in the outer
regions, it was not random in the more populated inner regions. This
suggests real variations of density at a given $R_{GC}$ and deserves further
study.


\section{Comparison with other types of objects}
Pagel et al. (1978) obtained oxygen abundances for HII regions spread over the LMC.
They found:
\begin{equation}
 12 + log (O/H) = -0.03 (\pm 0.02) \rho + 8.46 (\pm 0.06),
\end{equation}
over a range of galactocentric distance $\rho$ of 0 to 4 kpc.
Note that the parameters they used to derive $\rho$ are slightly
different from those used to obtain $R_{GC}$.
This will not make any qualitative difference to the result.
As Pagel et al. remark, a spatial gradient is evidently small
or absent in this young population.

Grocholski et al. (2006, 2007) have discussed metallicities of clusters in
the LMC. Very old clusters (SWB class VII) have low metallicities
([Fe/H]$\sim -1.7$) and are widely distributed over the LMC. Clusters of
intermediate age have a nearly constant metallicity: a mean [Fe/H] of --0.66
using metallicities corrected according to the prescription of C09. There is
no evidence of a gradient from $R_{GC} = 1$ to 6 kpc (see C09 fig.~2). The
intermediate age cluster NGC\,1718 in outer part of the LMC is an exception
to this with [Fe/H] =--0.8. However, there is some uncertainty regarding this
cluster (Grocholski et al. 2006 appendix A.1) If we adopt the C/M
calibration used by C09 (eq.6) together with our analysis (Fig.~5) the
population out to $R_{GC} \sim 4 kpc$ discussed in section 4 above, has a
distinctly lower mean metallicity ($[Fe/H] \sim -1.1$) than the intermediate
age clusters. In that case it presumably refers in the mean to a somewhat
older population. This age then falls in the well know LMC cluster age gap
where there are no known clusters with ages between $\sim 3$ and 10+ Gyr
(see for instance the summary by Da Costa 1991).

The mean metallicities of field RGB stars (e.g. Carrera et al. (2008)
are  more or less constant out to $R_{GC} \sim 6$ kpc and are similar
(on the scale of C09) to those of the intermediate age clusters but
with a lower abundance ([Fe/H] $\sim -0.8$ on the C09 scale) at $\sim 8$kpc.
There is also evidence of a smaller component with [Fe/H] $\sim -1.2$ 
(see C09
and references there).

\section{Conclusions} 
 Applying a period - [Fe/H] relation to the LMC RR Lyrae OGLE-III data base
we find that these stars show, in the mean, a small but significant radial
gradient in metallicity for distances out to beyond 5 kpc from the LMC
centre. RR Lyraes with spectroscopically estimated metallicities agree with
this result, though taken on their own the trend they show is only of marginal
significance. These results would evidently be consistent with the classical
picture of galactic evolution by the gradual collapse of a gas cloud.

A reanalysis of the data of C09 indicates that the C/M ratio of AGB stars
shows a very small change out to $\sim 4$kpc from the LMC centre and then
decreases dramatically. We attribute this decrease to an increasing
importance of an older population in the outer parts of the LMC. 
The very slight
increase in C/M from $R_{GC}$ = 1 to 4 kpc, if real, could be due to factors
such as age 
as well as metallicity and may also be
affected by uncertainties in assigning stars to the C or M classes on the
basis of photometric criteria.
The AGB population studied is apparently less metal rich
than the intermediate age clusters in the LMC, suggesting 
that in the mean this AGB population is older than the clusters
and has an age in the LMC cluster age gap.

\section*{Acknowledgments}
We are particularly grateful to Dr M-R. L. Cioni for her kindness in sending us
her extensive data on LMC AGB stars. Dr A. Layden very kindly
sent us his unpublished compilation of [Fe/H]  and period determinations for
Galactic RR Lyrae variables. This paper is based upon work supported
financially by the South African National Research Foundation.




\end{document}